\documentclass[reprint,aps,prd,floatfix,nofootinbibamsmath,amssymb,amsfonts]{revtex4-2}

\usepackage{bm}
\usepackage{graphicx,color}
\usepackage{physics}
\usepackage{dsfont}
\usepackage[normalem]{ulem}
\usepackage{hyperref,url}
\usepackage{mathrsfs}
\usepackage{calrsfs}
\usepackage{bbold}
\usepackage{mathtools}
\usepackage{comment}
\usepackage[caption=false]{subfig}
\usepackage{xcolor}

\usepackage{subfig}

\renewcommand{\vec}[1]{\mathbf{#1}}
\usepackage{varioref}
\usepackage[active]{srcltx}

\definecolor{purple1}{rgb}{128,0,128}

\expandafter\ifx\csname package@font\endcsname\relax\else
\expandafter\expandafter
\expandafter\usepackage
\expandafter\expandafter
\expandafter{\csname package@font\endcsname}%
\fi
\hypersetup{
	colorlinks=true,       
	linkcolor=blue,          
	citecolor=blue,        
}

\usepackage[section]{placeins}
\usepackage{enumitem}
\usepackage{fancybox}
\labelformat{figure}{Fig.~#1}

\DeclareMathAlphabet{\mathcal}{OMS}{cmsy}{m}{n}

\begin{document}
    \title{Expansion-contraction duality breaking in a Planck-scale sensitive cosmological quantum simulator} 

\author{S. Mahesh Chandran} 
\email{maheshchandran@snu.ac.kr}
\affiliation{Seoul National University, Department of Physics and Astronomy, Center for Theoretical Physics, Seoul 08826, Korea}
\author{Uwe R. Fischer} 
\email{uwerf@snu.ac.kr}
\affiliation{Seoul National University, Department of Physics and Astronomy, Center for Theoretical Physics, Seoul 08826, Korea}
%

\begin{abstract}
We propose the experimental simulation of cosmological perturbations governed by a Planck-scale induced Lorentz violating dispersion, aimed at distinguishing between early-universe models with similar power spectra. Employing a novel variant of the scaling approach for the evolution of a Bose-Einstein condensate with both contact and dipolar interactions, {we capture the hitherto unobserved phenomenon of trans-Planckian damping}. We show that scale invariance, and in turn, the duality of the power spectrum is subsequently broken at large momenta for an inflating gas, and at small momenta for a contracting gas. We thereby furnish a Planck-scale sensitive approach to analogue quantum cosmology that can readily be implemented in the quantum gas laboratory. 
\end{abstract}

\maketitle
\section{Introduction}




Inflation 
\cite{1981GuthPRD,1981SatoMNRAS,1982StarobinskyPLB,1982LindePLB,1983LindePLB,1982Albrecht.SteinhardtPRL} provides a causal mechanism for the generation of primordial density perturbations with a nearly scale-invariant power spectrum as observed in the Cosmic Microwave Background (CMB)~\cite{2003Bennett.etalAJS,2020Aghanim.etalAA}. However, a key limitation arises from the possibility that \textit{trans-Planckian} modes generated close to the initial singularity could also have redshifted to observable scales. 
 While a self-consistent treatment of such modes would require a UV-complete framework, 
 efforts to test the robustness of Hawking radiation against modified dispersions (arising e.g. in black-hole analogs~\cite{1981UnruhPRL}) inspired \textit{ad hoc} models of trans-Planckian physics~\cite{1991JacobsonPRD,1995UnruhPRD,1996Corley.JacobsonPRD,2005Unruh.SchutzholdPRD}. In the cosmological context, such models revealed that scale-invariance was generally not robust to short-distance modifications~\cite{2001Brandenberger.MartinMPLA,2001Martin.BrandenbergerPRD,2001NiemeyerPRD,2001Niemeyer.ParentaniPRD,2001StarobinskyJEaTPL,2002Niemeyer.etalPRD,2002Brandenberger.etalPRD,2002DanielssonPRD,2003Burgess.etalJHEP,2003ShankaranarayananCQG,2013Brandenberger.MartinCQG} -- tightly constraining the 
assumptions on trans-Planckian physics in inflationary scenarios~\cite{2001Easther.etalPRD,2002Kaloper.etalPRD,2020Bedroya.etalPRD,2020BrahmaPRD,2025Brahma.CalderonFigueroa}.  

\begin{figure}[!b]
	\centering
	\includegraphics[scale=0.55]{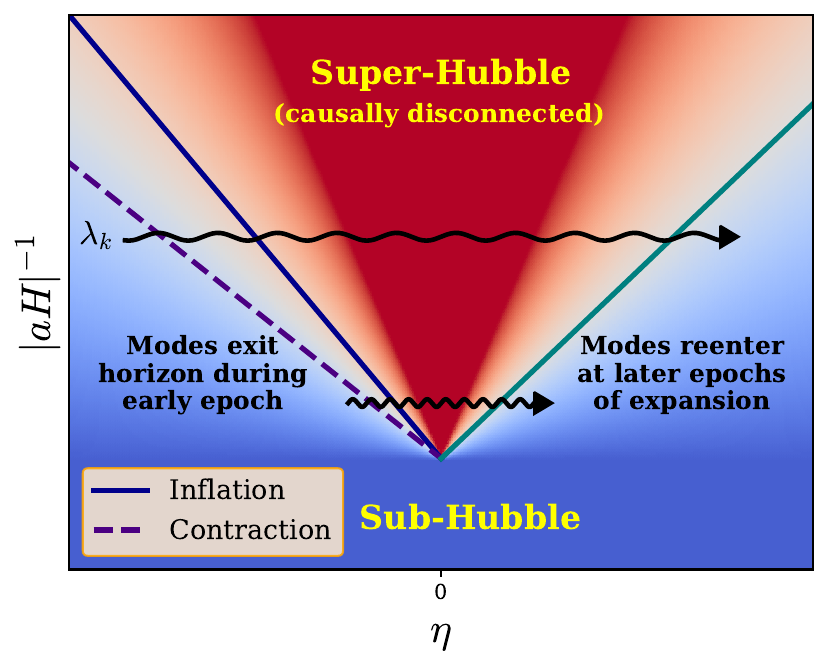}
	\caption{Timeline of comoving horizon $|aH|^{-1}=|\frac{\eta}{v}|$ (corresponding to cosmological 
	scale factor $a\propto |\eta|^v$), and comoving 
    mode propagation prior to reaching current observable CMB scales. A scale-invariant power spectrum can be generated in the early epoch ($\eta<0$), either via inflation ($v=-1$), or via a \textit{contraction} phase ($v=3$) leading up to bounce (set at $\eta=0$).}
	\label{fig:ModeCrossing}
\end{figure}

The idea of a bouncing cosmology~\cite{2008Novello.BergliaffaPR} 
can circumvent 
the initial singularity and prevent trans-Planckian modes from reaching observable scales: An initial contraction phase generates 
primordial density perturbations (as schematically illustrated in \ref{fig:ModeCrossing}) with a scale-invariant power spectrum similar to inflation via  a \textit{duality invariance} of perturbation spectra corresponding to certain expanding and contracting backgrounds~\cite{1999WandsPRD,2002Finelli.BrandenbergerPRD,2012Brandenberger}. 
Since both inflation \cite{2004SteinhardtMPLA,2013Ijjas.etalPLB} 
and bouncing \cite{2005Cattoen.VisserCQG,2015Battefeld.PeterPR,2017Nojiri.etalPR}
models encounter unresolved conceptual (as well as fine-tuning) issues, this duality in fact 
{\em weakens} the 
power spectrum as a unique indicator of early-universe possibilities~\cite{2024Raveendran.ChakrabortyGRG,2024Chandran.ShankaranarayananIJMPD}.



 Given the inherent challenges in recreating the initial conditions of the Universe, cosmology has largely relied on observations. To enable addressing cosmological issues via reproducible experiments, 
 the era of {\em analogue quantum cosmology} using 
ultracold gases was ushered in, first theoretically
~\cite{CarlosBarcelo_2001,2003Barcelo.etalIJMPD,2003Barcelo.etalPRA,PhysRevLett.91.240407,FedichevPRA2004,PhysRevA.70.063615}, 
and then
culminating in pioneering experiments 
that {observed various cosmological phenomena} 
\cite{2013Hung.etalS,Eckel,Banik,2022Viermann.etalNature,Tajik}. For reviews 
of what has more generally, 
covering a broad range of  
physical systems, been dubbed {\em analogue gravity},  see 
\cite{Barcelo:2005fc,2023Braunstein.etalNRP,SCHUTZHOLD2025104198}. 
{This has led to major milestones, inter alia}, 
the first observation of the quantum Hawking radiation effect in 
a Bose-Einstein condensate (BEC) \cite{Steinhauer:2015saa, MunozdeNova:2018fxv,Kolobov2021}. 
These systems 
also 
allow, 
in principle, the observation of the more elusive quantum Unruh effect in its various guises \cite{Chin_Unruh,Gooding}.
While contact interactions between the gas constituents have remained the primary focus, 
recent studies taking into account dipole-dipole interactions~
\cite{Chomaz} have significantly enriched 
BEC simulations, particularly for exploring the impact of 
trans-Planckian dispersion 
\cite{2017Chae.FischerPRL,TianPRD,PhysRevD.107.L121502}.

In what follows, we propose 
the experimental simulation of primordial density perturbations in a quasi-2D dipolar BEC, 
wherein a strongly confining transverse trap introduces an effective Planck-scale that dynamically alters the standard Lorentz-invariant dispersion. We show that a novel anisotropic variant of the usual scaling approach~\cite{CastinDum,Kagan} 
 can be used to engineer a dispersion of the form $\omega_k\propto a\lambda_{pl}^{-1}F[k\lambda_{pl}/a]$, previously employed to 
address trans-Planckian issues in cosmology~\cite{2001Brandenberger.MartinMPLA,2001Martin.BrandenbergerPRD,2001NiemeyerPRD,2001Niemeyer.ParentaniPRD,2001StarobinskyJEaTPL,2002Niemeyer.etalPRD,2002Brandenberger.etalPRD,2002DanielssonPRD,2003Burgess.etalJHEP,2003ShankaranarayananCQG,2013Brandenberger.MartinCQG}, 
and which arises in UV-complete quantum gravity candidates (such as Horava-Lifshitz~\cite{2009HoravaPRD,2010MukohyamaCQG,2012Ferreira.BrandenbergerPRD}) 
as well as string-theory motivated minimal-length models~\cite{2001KempfPRD,2001Kempf.NiemeyerPRD,2003Hassan.SlothNPB}. 
It is explicitly demonstrated that the duality invariance of the power spectrum corresponding to expanding--contracting cosmologies~\cite{1999WandsPRD} is broken by {the effects of trans-Planckian 
damping}~\cite{2003Hassan.SlothNPB}
, as 
can be experimentally verified in our quantum simulator. Specifically, 
this damping tilts the spectrum at 
large momenta for inflation and at 
small momenta for contraction.
 Aided by these observations, we discuss how ultracold-atom experiments can isolate potential \textit{large-scale} signatures of 
Lorentz violation in the power spectrum, 
and distinguish between competing early-universe models. 

We further 
highlight a microscopically controlled variant of the dispersion first 
put forward by Unruh~\cite{1995UnruhPRD}.
This ``quasi-flat band" (nearly zero group velocity over a large range of wavevectors) 
dispersion is shown to 
{promote trans-Planckian damping to a leading order effect, resulting }
in the {\em freezing} (akin to inflation) of an otherwise growing contraction power spectrum along with a slight red-tilt similar to current CMB observations~\cite{2020Akrami.etalAA}. 
 Remarkably, the Unruh dispersion corresponds to exactly {\em equal}  
dipolar ($g_d$) and contact ($g_c$) couplings, which coincides with the stability boundary of the bulk quantum gas in the embedding three spatial dimensions of our 2+1D setup, 
cf.~Refs.~\cite{2006FischerPRA,2008Koch.etalNatPhys}.
The Unruh dispersion is, thus, readily experimentally realizable in dipolar-contact BECs with only moderate Feshbach tuning of the contact interaction necessary, for example in Dysprosium and Erbium condensates with their large magnetic dipole moments of $10$ and $7\,\mu_B$, respectively \cite{Lev,Ferlaino}. Hence Planck-scale sensitive analogue cosmology can be realized with current experiments in the quantum gas laboratory.

\section{Setup}  
We consider a 3D quantum gas subject to contact as well as dipole-dipole interactions, with the dipoles aligned along the $z$ direction.  We confine the gas 
to a  harmonic oscillator ground state in the 
transverse direction \cite{2006FischerPRA} with an 
evolving trap frequency 
$\omega_z(t)$ that scales the oscillator length as $d_z(t)=b_z(t)d_{z,0}$
. {For a tight axial trap width ($\omega_z\gg \omega_x,\omega_y$), the $z$-component can be integrated out to obtain an effective} quasi-2D condensate. We then employ the scaling approach --- an established procedure for analyzing BECs with
generally time dependent coupling strengths placed in time-varying traps~\cite{CastinDum,Kagan}. 
However, we relax a central assumption in the otherwise general scaling approach of \cite{2010Gritsev.etalNJP}, namely that the 
time dependence of pairwise interaction terms can be collected as 
$V(\vec{r};t)= \mathscr{V}(t)V(\vec{r})$. Distinct from earlier works, we can 
therefore address a form of {\em anisotropic} 
scaling along radial and transverse directions tailored to our simulation purposes, which results in a time dependent modified dispersion in the comoving frame of the quasi-2D condensate (for details, 
see Appendix \ref{Sec. I}.  

In the comoving frame defined by the 2D scaled coordinate $\vec{x}=\vec{r}/b(t)$, the fluid density is approximately constant ($\rho\sim\rho_0$), and the velocity vanishes ($\partial_t\vec{x}\sim 0$). 
By synchronizing the transverse scaling with respect to the desired cosmological scale factor as $b_z=a^2$ (which allows the contact and dipolar coupling strengths, i.e., $g_{c,0}$ and $g_{d,0}$ respectively, to be kept constant 
(Appendix \ref{Sec. I}), and linearizing the fluctuations on top of 
the condensate 
phase ($\phi_0+\delta\phi$) and density ($\rho_0+\delta\rho$),
we get the corresponding equations of motion in the momentum space as follows: 
\begin{align}
    \delta\ddot{\phi}_k&+\left(2\frac{\dot{a}}{a}-\frac{\dot{W}_k}{W_k}\right)\delta\dot{\phi}_k+\frac{c_0^2k^2W_k}{a^2}\delta\phi_k=0,\nonumber\\
    \delta\ddot{\rho}_k&+\frac{c_0^2k^2W_k}{a^2}\delta\rho_k=0,
\end{align}
where we have defined, cf.~\cite{2006FischerPRA}
\begin{align}
    W_k&=1-\frac{3R}{2}w\left[\frac{b_zkd_{z,0}}{b}\right]+\frac{k^2d_{z,0}^2a^2}{4A};\\
    R&=\frac{g_{d,0}}{d_{z,0}g_{\rm eff,0}};\quad A=\frac{mc_0^2}{\hbar \omega_{z,0}};\quad  c_0^2=\frac{g_{\rm eff,0}\rho_0}{m},\nonumber
\end{align}
with an effective contact coupling 
 $g_{\rm eff,0}=(g_{c,0}+2g_{d,0})/\sqrt{2\pi}d_{z,0}$, and where  $w[z]= ze^{\frac{z^2}{2}}(1-\erf(z/\sqrt{2}))$. The derivatives ($\dot{f}=\partial_{\tau}f$) above are with respect to the scaling time $\tau=\int b^{-2}dt$. In the long-wavelength limit of $W_k\to1$, the phase fluctuation dynamics exactly maps to that of 
 a massless, minimally coupled scalar field propagating in a (2+1)-dimensional Friedmann-Lema{\^i}tre-Robertson-Walker space-time~\cite{1982Birrell.Davies,2005Jacobson}:
\begin{align}
    \square\delta\phi&=\frac{1}{\sqrt{|g|}{}} \partial_\mu\left(\sqrt{|g|}g^{\mu \nu} \partial_\nu \delta \phi\right)=0;\nonumber\\
    ds^2&=g_{\mu\nu}dx^\mu dx^\nu=d\tau^2-a^2(\tau)d\vec{x}^2.
\end{align}
The second and third terms in $W_k$ correspond to dipolar interaction and free-particle contribution respectively, which 
break Lorentz invariance at shorter wavelengths.

\begin{figure}[t]
	\begin{center}
		\subfloat[\label{PC1}][Trans-Planckian dispersion]{%
			\includegraphics[width=0.24\textwidth]{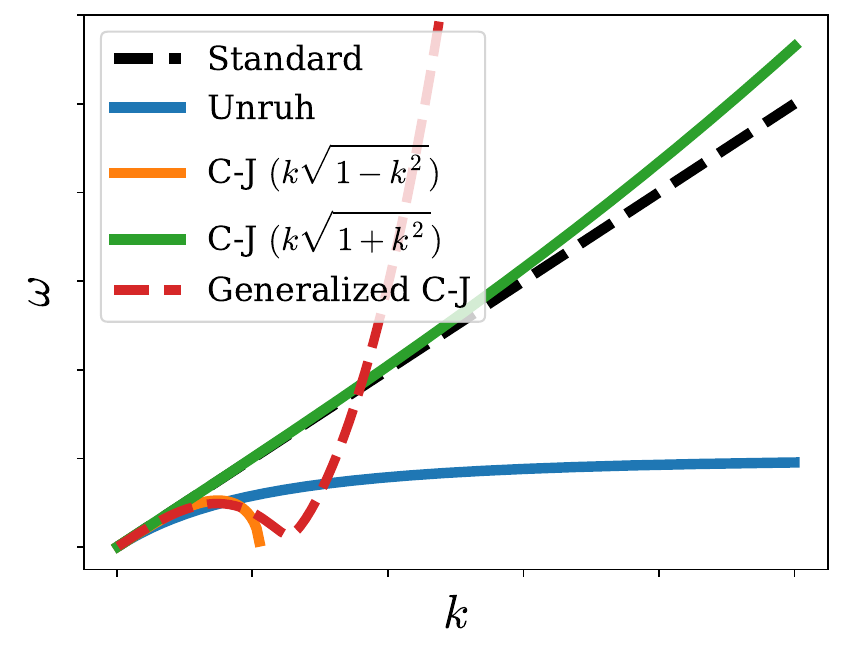}
		}
		\hspace*{-0.5em}
		\subfloat[\label{PC2}][Dipolar BEC dispersion]{%
			\includegraphics[width=0.24\textwidth]{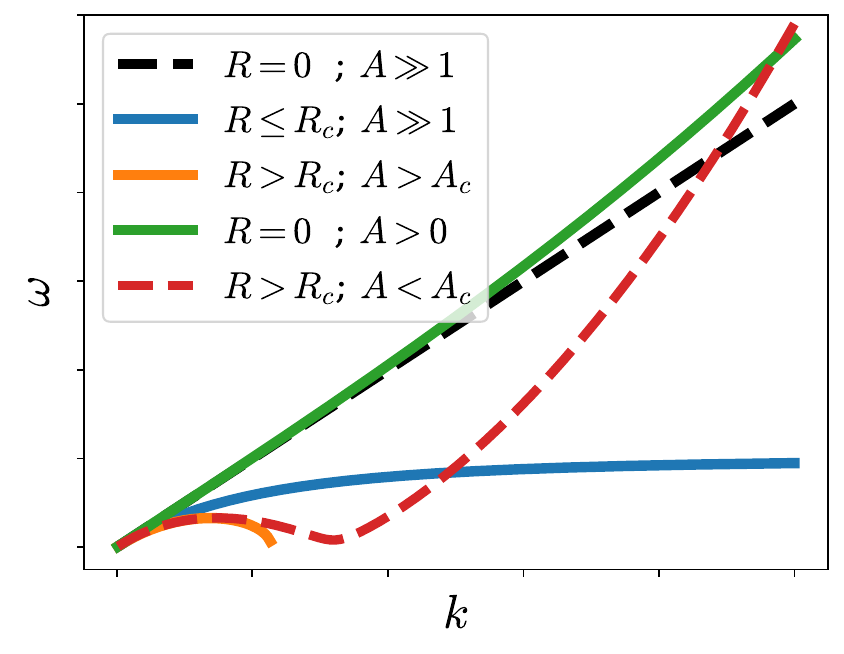}
		}
		
		\caption{Side-by-side comparison of dispersions corresponding to (a) well-known trans-Planckian models, and (b) quasi-2D dipolar BECs. One can simulate subluminal (e.g., Unruh~\cite{1995UnruhPRD}) or superluminal (e.g., C-J for Corley-Jacobson~\cite{1996Corley.JacobsonPRD}) cases by tuning $R$ and $A$. Dispersions with a minimum (such as those appearing in generalizations of the C-J type~\cite{2002Brandenberger.MartinIJMPA}) can also be modeled via roton minimum tuning (dashed red line)~\cite{2006FischerPRA,2017Chae.FischerPRL}.}
		\label{fig:disp}
	\end{center}
\end{figure}

By appropriately tuning the parameters $R$ (relative dipolar strength) and $A$ (dimensionless sound speed), one can simulate a variety of nonlinear trans-Planckian 
dispersions,  as shown in \ref{fig:disp}. However, our aim here is to probe Planck-scale effects in cosmology via a dispersion of the form $k^2W_k=a^2\lambda_{pl}^{-2}F^2(k\lambda_{pl}/a)$~\cite{2001Brandenberger.MartinMPLA,2001Martin.BrandenbergerPRD,2001NiemeyerPRD,2001Niemeyer.ParentaniPRD,2001StarobinskyJEaTPL,2002Niemeyer.etalPRD,2002Brandenberger.etalPRD,2002DanielssonPRD,2003Burgess.etalJHEP,2003ShankaranarayananCQG,2013Brandenberger.MartinCQG} --- wherein high-momentum modes fall back to the standard Lorentz-invariant dispersion after being redshifted to sub-Planckian scales by the expansion. 
{ We can realize this cosmologically relevant 
dependence on $k\lambda_{pl}/a$ in the BEC dispersion by (i) 
employing 
a novel type of anisotropic scaling
along radial and transverse directions such that $b=ab_z$, in addition to (ii) ignoring the free-particle contribution ($k^{2}d_{z,0}^2a^2\ll 4A$):
    \begin{equation}\label{eq:scaling2}
         W_k\approx 1-\frac{3R}{2}w\left[\frac{kd_{z,0}}{a}\right],
    \end{equation}
where the Planck length $\lambda_{pl}$ for our effective 2+1D space-time is set by the trap-width $d_{z,0}$ along the compactified extra (transverse) dimension}. We confine the relative strength $R$ up to a critical value $R_c\coloneqq \sqrt{2\pi}/3$, at which the dispersion coincides almost exactly with that of Unruh~\cite{1995UnruhPRD}, $k^2W_k\approx a^2d_{z,0}^{-2}\tanh^{2/p}{\left[(kd_{z,0}/a)^p\right]}|_{p\to1/2}$, asymptoting to zero group velocity at large momentum (\ref{fig:disp}). {Note that this form of dispersion is naturally preferred in minimal length models of cosmology that impose a cutoff on the transformed physical momenta as $kW_k^{1/2}/a\leq \lambda_{pl}^{-1}$~\cite{2003Hassan.SlothNPB}}. Beyond the critical strength $R_c$, the dispersion becomes unstable at large momentum, unless stabilized by the free-particle term (resulting in a \textit{roton minimum} \cite{Chomaz}). Though this free-particle term is of the Corley-Jacobson type (\ref{fig:disp})~\cite{1996Corley.JacobsonPRD}, it \textit{amplifies} 
Lorentz violation 
with the redshifting of modes, {i.e., it has no cosmological analogue}. It is thus ideal to confine the experimental protocol to time- and momentum-scales where this term can be safely ignored, beyond which the mapping to $2+1$-D cosmological perturbation theory 
{incurs free-particle corrections} (see Appendix \ref{Sec. II}) 
The limit of 
{$k^2d_{z,0}^2a^2 \ll 4A$} henceforth assumed (keeping $R\le R_c$) 
then sufficiently captures low-momentum signatures 
{of direct relevance to cosmology}. 

\section{Fluctuation power spectrum}  Below we use units $\hbar=m=1$, and also set $c_0=1$ and $d_{z,0}=1$ (our 
Planck scale in the extra dimension). {For a scale factor that evolves as a power-law ($v$) in terms of conformal time $\eta$:
\begin{equation}
    a(\eta)=\left(\frac{\eta}{\eta_i}\right)^v;\quad \eta=\int \frac{d\tau}{a};\quad \phi_k'=\partial_\eta\phi_k,
\end{equation}}
we get the following mode evolution equation in terms of the rescaled field variable $\delta\bar{\phi}_k=\sqrt{a/W_k}\delta\phi_k$:
\begin{align}\label{eq:disp}  &\delta\bar{\phi}''_k+\omega_k^2\delta\bar{\phi}_k=0;\quad \omega_k^2=k^2W_k+\frac{v(2-v)}{4\eta^2}\left(1+\Delta_k\right);\nonumber\\
    &
    \Delta_k=\frac{(4v-2)a\partial_aW_k+2va^2\partial_a^2W_k}{(2-v)W_k}-\frac{3va^2(\partial_aW_k)^2}{(2-v)W_k^2}.
\end{align}
{The dispersion features two key modifications --- Lorentz-invariance is dynamically violated by the $k^2W_k$ term (as conjectured in ad hoc approaches to cosmological trans-Planckian physics~\cite{2001Brandenberger.MartinMPLA,2001Martin.BrandenbergerPRD,2001NiemeyerPRD,2001Niemeyer.ParentaniPRD}), and a nonadiabatic correction $\Delta_k$ is generated by the time-dependence of $W_k$. While ad hoc models have generally failed to incorporate the latter (i.e., $\Delta_k$), it is an inherent feature of minimal-length frameworks that pertains to the phenomenon of \textit{trans-Planckian damping}~\cite{2003Hassan.SlothNPB}. By capturing this feature, the BEC analogue can provide novel insights (as discussed below) and also guide the improvement of ad hoc models where a full quantum gravity treatment is lacking. Its scope may be further expanded by engineering a source term in the evolution equation~\cite{2005Shankaranarayanan.LuboPRD,2012Ferreira.BrandenbergerPRD}, facilitating the simulation of a wider variety of quantum gravitational effects that extend into the semiclassical domain.} 

To first understand the $\Delta_k\sim 0$ case, let us consider a nearly static $W_k$ which can be realized in the lab via isotropic scaling ($b_z=b$) of the dipolar condensate~\cite{2017Chae.FischerPRL}, away from the free-particle regime~({$k^2d_{z,0}^2a^2 \ll 4A$}). The mode functions evolving from the (minimum energy~\cite{2001Brandenberger.MartinMPLA,1978Brown.DuttonPRD}) vacuum state defined at $\eta\to-\infty$ are given below:
\begin{equation}
    \delta\bar{\phi}_k=\frac{\sqrt{\pi\abs{\eta}}}{2}H_s^{(1)}\left(\omega_k^{\rm in}|\eta|\right);\quad \lim_{\eta\to -\infty}\delta\bar{\phi}_k\to\frac{e^{-i\omega_k^{\rm in}\eta}}{\sqrt{2\omega_k^{\rm in}}},
\end{equation}
where $\omega_k^{\rm in}=k\sqrt{W_k^{\rm}}$ corresponds to the initial $k$-mode frequency, and $s=\abs{v-1}/2$ is the
Hankel function index. Each mode evolves from sub-Hubble to super-Hubble scales, with curvature effects taking over as they cross the horizon at $k\eta=-1$. The power spectrum at superhorizon scales ($\abs{k\eta}\ll 1$) takes the form:
\begin{equation}\label{eq:SIPS}
    \mathcal{P}_{\delta\phi}\coloneqq k^2\abs{\delta\phi_k}^2\simeq  \abs{\frac{H}{v\pi}}\left(\frac{4}{k^2W_k\abs{\eta}^2}\right)^{s-1}\Gamma^2(s).
\end{equation}
For the special case $s=1$, the spectrum is scale invariant. This is the 2D counterpart of the standard Lorentz-invariant result in 3D~\cite{1999WandsPRD}, which here has further been generalized to an adiabatic modification to the dispersion. A useful measure for the power spectrum is the scalar spectral index $n_s-1=d(\ln \mathcal{P}_{\delta \phi})/d\ln{k}$. Scale-invariance corresponds to $n_s=1$ (adiabatic case considered above), a blue spectral tilt when $n_s>1$ (more clumpiness at shorter length scales) and a red spectral tilt when $n_s<1$ (more clumpiness at larger length scales). 

\section{Duality invariance} 
Cosmological models related by the transformation $v\to 2-v$ exhibit \textit{duality invariance}, wherein the dispersion \eqref{eq:disp} remains invariant and the perturbation spectra share 
the same scale dependence~\cite{1999WandsPRD,2002Finelli.BrandenbergerPRD}. 
For instance, the scale-invariant case $s=1$ corresponds to the following dispersion (when $\Delta_k\sim0$):
\begin{equation}\label{eq:dualdisp}
        \omega_k^2= k^2W_k-\frac{3}{4\eta^2},
    \end{equation}
which can be attributed to either a de Sitter expansion ($v=-1$), or a \textit{contraction} scenario ($v=3$) characterized by an equation of state $w\coloneqq P/\rho=1/3$ that lies between a radiation dominated ($w=1/2$) and a matter dominated ($w=0$) background in 2+1-D~\cite{2010Letelier.PitelliPRD} (in 3+1, the dual to de Sitter is a matter-dominated contraction~\cite{1999WandsPRD}). These models are dual in the sense that they both generate a power spectrum that is scale invariant at superhorizon scales, with the only difference being that $\mathcal{P}_{\delta\phi}$  
freezes for de Sitter ($\abs{H}$ is constant) and grows for contraction (with growing  $\abs{H}$). 

\begin{figure}[b]
	\begin{center}
		\subfloat[][$R=0$]{%
			\includegraphics[width=0.24\textwidth]{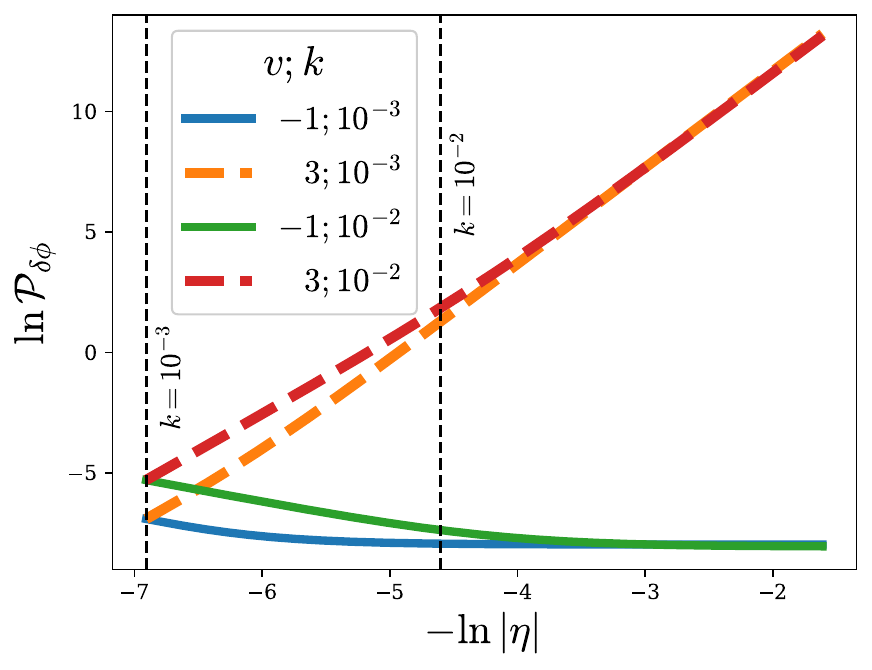}
		\label{fig:PST1a}}
		\subfloat[][$R=0.4$]{%
			\includegraphics[width=0.24\textwidth]{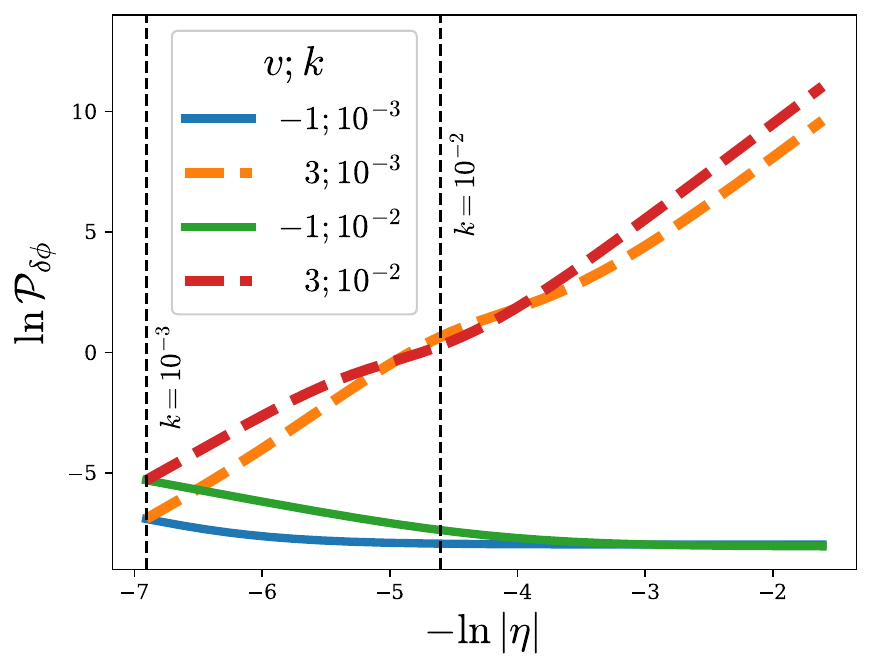}\label{fig:PST1b}
		}\\
        \subfloat[][$R=0.8$]{%
			\includegraphics[width=0.24\textwidth]{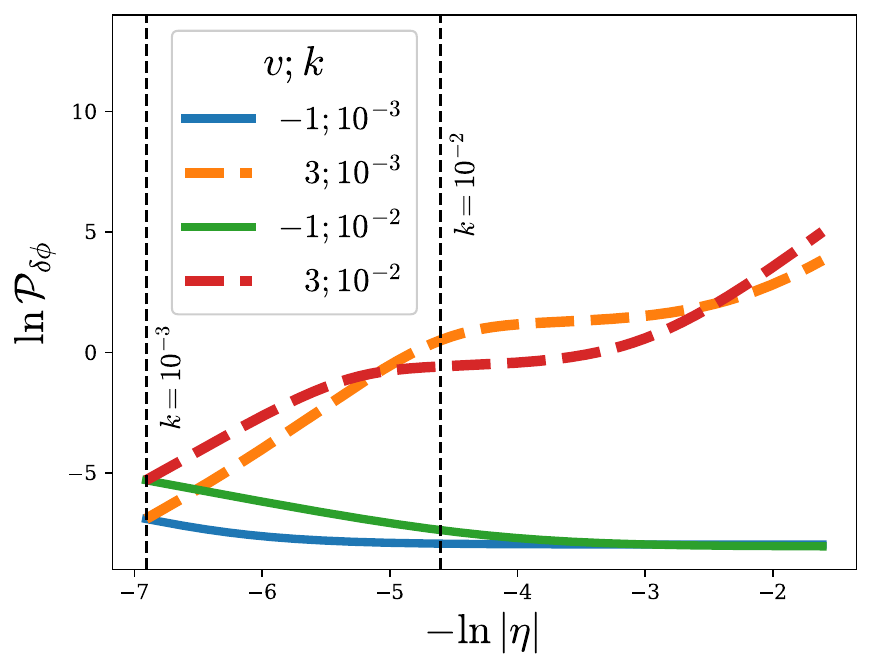}
		\label{fig:PST1c}}
		\subfloat[][$R=R_c$]{%
			\includegraphics[width=0.24\textwidth]{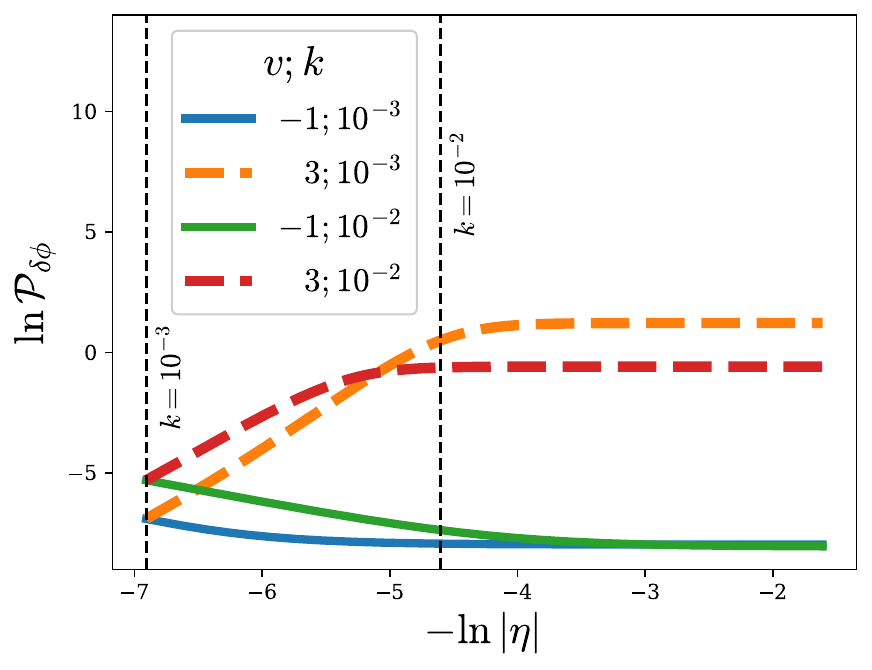}\label{fig:PST1d}
		}
		\caption{Time evolution of inflation ($v=-1$) and contraction ($v=3$) power spectra $\mathcal{P}_{\delta\phi}$ for various values of relative dipolar strength $R$. The black vertical lines indicate the horizon crossing times ($|k\eta|=1$) corresponding to the low-momentum ($k=10^{-3},10^{-2}$) modes considered here. The scale-invariance duality is preserved for $R=0$, and broken for $R>0$.}
		\label{fig:PST}
	\end{center}
\end{figure}

In the context of the scaling approach, duality implies that a scale-invariant power spectrum can be simulated by a dipolar Bose-gas that is either expanding exponentially ($a\propto e^{H\tau}$) or contracting as a power law ($a\propto (-\tau)^{3/4}$). The latter offers a convenient alternative for reproducing scale invariance in the laboratory due to the following reasons --- (i) the free-particle crossover that cuts off the analogue-cosmology mapping during expansion is avoided as contraction proceeds, and (ii) density-contrast measurements~\cite{2022Viermann.etalNature} can achieve better temporal resolution for a power-law evolution over an exponentially-fast evolution, allowing a more accurate extraction of the phase-fluctuation spectrum. As shown in \eqref{eq:SIPS}, the duality is also robust to a generalized dispersion as long as $W_k$ is nearly static, with scale invariance persisting despite broken Lorentz-invariance.



\section{Duality breaking} To simulate trans-Planckian dispersion in the early-universe, the modification $W_k$ must be dynamically driven by dipole-dipole interactions subject to an anisotropic scaling ($b=ab_z$) of the gas along transverse and radial directions~\eqref{eq:scaling2}. Consequently, the emergence of {\em trans-Planckian damping} via $\Delta_k$ breaks the duality between inflation and contraction due to its sensitivity to the power law (it is no longer invariant under $v\to2-v$). This leads to a modified late time ($\eta\to0^-$) dispersion as follows:
     \begin{equation}\label{eq:correction}
       \omega_k^2 \simeq 
        \begin{cases}  k^2W_k-\frac{3}{4\eta^2}\left(1-\frac{Rk}{a}\right) &  v=-1\\ k^2W_k-\frac{3}{4\eta^2}\left(1-\frac{32Ra^2}{(R_c-R)k^2}\right) &  v=3 \end{cases},
    \end{equation}
where the duality breaking at superhorizon scales becomes apparent, with the leading order corrections to the adiabatic case \eqref{eq:dualdisp} (and in turn, to scale-invariance) being prominent at large $k$ for inflation and small $k$ for contraction. While the effect is amplified with increasing $R$ in both cases, contraction leads to a different late-time behaviour very close to the critical ``Unruh" value $R\to R_c$ 
(note that the $a\to0$ and $R\to R_c$ limits of $\Delta_k$ as defined in the second line of 
\eqref{eq:disp} do not commute):
\begin{equation}\label{eq:critdisp}
       \omega_k^2 \simeq  k^2W_k-\frac{3}{4\eta^2}\left(5+\frac{24a^2}{k^2}\right).
    \end{equation}
    {Here, the damping term $\Delta_k$ is promoted to leading order, and dominated by a} scale-independent term absent in the noncritical case~\eqref{eq:correction}. 
    In the leading order, this dispersion now matches that of a $v=-3$ power-law expansion, which for 
    a nearly static $W_k$ results in a frozen power spectrum $\mathcal{P}_{\delta\phi}\propto (k^2W_k)^{-1}$ that is scale invariant ($n_s= 1$) at large $k$ and red-tilted ($n_s<1$) at small $k$. {Importantly, we can conclude that these results also carry over to minimal length approaches, revealing hitherto unexplored, dramatic implications of trans-Planckian damping~\cite{2003Hassan.SlothNPB} 
    on collapsing geometries (see for details Appendix \ref{Sec. III})}. 


\begin{figure}[t]
\vspace*{1em}
	\begin{center}
		\centering
	\includegraphics[scale=0.5]{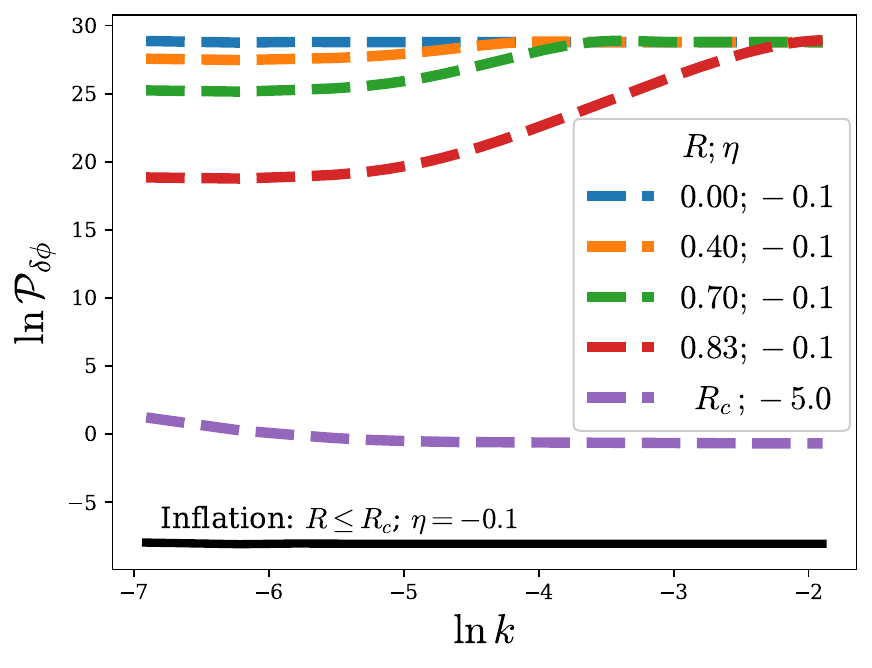}
				\caption{Scale dependence of the 
                superhorizon power spectrum $P_{\delta\phi}$ corresponding to low-momentum ($k\ll1$) modes for inflation (black line) and contraction (dashed lines). Lorentz violation tilts the spectrum at small $k$ exclusively for contraction; to a blue tilt for $0<R<R_c$ and a red tilt at $R_c\sim0.835$.}
		\label{fig:PS2}
	\end{center}
\end{figure}

In order to understand how these corrections translate to observable consequences, we rely on numerical simulations of the power spectrum (incorporating the exact form of $\Delta_k$, see for details of the simulations Appendix \ref{Sec. IV},  
and track its evolution all the way to superhorizon scales $|k\eta|\ll1$. While \eqref{eq:correction} indicates duality breaking due to nonadiabatic corrections at both the low-momentum corner (for contraction) and high-momentum corner (for inflation), our focus is on the former. From \ref{fig:PST} and \ref{fig:PS2}, we observe that in this regime~($k\ll1$) the inflationary power spectrum remains scale invariant, whereas nonadiabatic corrections to the contraction power spectrum manifest as a blue-tilt ($n_s>1$) for noncritical values of $R$, and a red-tilt ($n_s<1$) for $R=R_c$. At the critical ``Unruh" value $R_c$, the contraction power spectrum freezes (as seen in \ref{fig:PST}) due to the late-time dynamics asymptoting to that of a $v=-3$ power-law expansion~\eqref{eq:critdisp} to leading order.  

\section{Conclusion} 
{We have proposed a novel experiment for the simulation of trans-Planckian models previously conjectured in early-universe cosmology, and the isolation of potential \textit{low-energy} imprints encoded in the power spectrum}. For adiabatic nonlinear dispersion ($\Delta_k\sim0$), simulated via conventional isotropic scaling of dipolar condensates ($b=b_z$), the scale-invariance duality between inflation and contraction spectrum is preserved. When however, nonadiabatic corrections arising from a Planck-scale sensitive dispersion in our novel anisotropic scaling setup ($b=ab_z$) occur, the duality is broken. Whereas the inflationary power spectrum washes out any low-momentum signature of Planck-scale physics with the expansion, contraction clearly records these signatures via a spectral tilt in the low-momentum corner --- implying that although trans-Planckian modes may never reach observable scales in a bouncing model, Lorentz violation can still leave imprints at \textit{large scales} from the initial contraction phase. {Though such imprints are rarely discussed in the  literature~\cite{2004Shankaranarayanan.SriramkumarPRD,2002Brandenberger.HoPRD}, they have a clear origin here by virtue of trans-Planckian damping, which happens to be an essential feature of minimal length models in cosmology~\cite{2001KempfPRD,2001Kempf.NiemeyerPRD,2003Hassan.SlothNPB}. For the critical ``Unruh" case $R=R_c$, this damping is promoted to a leading order effect, resulting in a \textit{frozen} contraction power spectrum (akin to inflation) with a slight red-tilt similar to current CMB observations~\cite{2020Akrami.etalAA}. Since these results also carry over to more general models that incorporate such damping terms~(see part \ref{Sec. III} of Appendix), the BEC setup captures previously unexplored signatures that are quite robust to the underlying theory, with profound implications for contracting backgrounds in particular. The possibility of further expanding this setup via an entropic source term~\cite{2005Shankaranarayanan.LuboPRD,2012Ferreira.BrandenbergerPRD} will be discussed elsewhere --- with the goal of simulating a wider variety of quantum gravitational effects that extend into the semiclassical regime. 
{Additionally, since the validity of the analogue cosmology mapping is seemingly enhanced in the 3D crossover regime (i.e., $A\gg1$), a more faithful treatment incorporating mean-field effects from the transverse direction is required to further probe this regime, which we leave for future work.}

{By furnishing a novel scaling approach for dipolar BECs that faithfully simulates Planck-scale physics in the early-universe, we have bridged a long-standing gap in realizing experimental cosmology with ultracold quantum gases}. As a precise application, we have demonstrated the utility of our analogue model 
in isolating observable, low-energy effects that can distinguish between competing early-universe models and aid our interpretation of current day observations.} Finally, our findings can be validated in currently realized magnetically dipolar BECs, with only moderate Feshbach tuning of the contact interaction being applied, due to the fact that we stay outside the roton regime throughout.

\acknowledgments

This work has been supported by the National Research Foundation of Korea under 
Grant No.~2020R1A2C2008103.

\begin{appendix}

\begin{widetext}



\section{Setting up a Planck-scale sensitive cosmological quantum simulator}
\label{Sec. I}

A collective description of atoms or molecules of mass $m$ in a Bose gas is captured by the following Lagrangian density~\cite{2017Chae.FischerPRL}:
\begin{equation}
    \mathcal{L}=\frac{i\hbar}{2}\left(\Psi^*\partial_t \Psi-\Psi\partial_t \Psi^*\right)-\frac{\hbar^2}{2m}|
    \nabla\Psi|^2-V_{\rm ext}|\Psi|^2-\frac{1}{2}|\Psi|^2\int d^3 \vec{R'} V_{\rm int}(\vec{R}-\vec{R'})|\Psi(\vec{R'})|^2,
\end{equation}
where $\vec{R}=(\vec{r},z)$ are spatial 3D-coodinates and $V_{\rm ext}\coloneqq m\omega^2r^2/2+m\omega_z^2z^2/2$ is the trapping potential with frequencies that are generally time dependent. The interaction term $V_{\rm int}(\vec{R})=g_c\delta^3(\vec{R})+V_{\rm dd}(\vec{R})$ is characterized by the contact interaction coupling ($g_c$) as well as $V_{\rm dd}(\vec{R})=(3g_d/4\pi)(1-3z^2/|\vec{R}|^2)/|\vec{R}|^3$ corresponding to dipoles polarized perpendicular to the plane. We confine ourselves to the quasi-2D regime by keeping the gas tightly compact along the transverse direction over the course of the expansion. To effectively model this system, we decompose the field along radial and transverse direction as $\Psi=\Psi_r\Phi_z$, and assume that the transverse component is described by a ground state harmonic oscillator wavefunction corresponding to a time dependent trapping frequency~\cite{2008LoheJPA,2017Chae.FischerPRL}:
\begin{align}\label{eq:Ermakov1}
    \Phi_z(z,t)&=\left(\frac{1}{\pi d_z^2}\right)^{\frac{1}{4}}\exp\left[-\frac{z^2}{2d_z^2}+\frac{im\dot{b}_zz^2}{2\hbar b_z}-\frac{i\omega_{z,0}}{2}\int\frac{dt}{b_z^2}\right];\nonumber\\
    d_z(t)&=b_z(t)d_{z,0};\quad d_{z,0}=\sqrt{\frac{\hbar}{m\omega_{z,0}}};\quad \partial_t^2 b_z+\omega_z^2b_z=\frac{\omega_{z,0}^2}{b_z^3},
\end{align}
where $b_z$ is the scaling parameter that solves the nonlinear Ermakov-Pinney equation~\cite{1950PinneyPotAMS} corresponding to a time dependent frequency $\omega_z(t)$. {Note that the Ermakov equation is valid in the quasi-2D regime where the axial kinetic energy dominates over interactions ($\hbar\omega_{z,0}\gg mc_0^2$). Though a more general treatment of this ansatz may be possible via the variational approach~\cite{2006FischerPRA}, we confine to the quasi-2D regime to facilitate a straightforward mapping to cosmological perturbations.} Integrating out the transverse component, we get a dimensionally reduced Lagrangian:
\begin{equation}
    \mathcal{L}_r=\frac{i\hbar}{2}\left(\Psi_r^*\partial_t{\Psi}_r-\Psi_r\partial_t{\Psi}^*_r\right)-\frac{\hbar^2}{2m}|\nabla_r\Psi_r|^2-\frac{1}{2}m\omega^2r^2|\Psi_r|^2-\frac{1}{2}\int d^2r' V_{\rm int}^{2D}(\vec{r}-\vec{r}')|\Psi_{r'}|^2|\Psi_r|^2.
\end{equation}
We now shift to the comoving frame $\vec{x}$ of the quasi-2D condensate, corresponding to its expansion/contraction along the radial direction by a scale factor $b(t)$. For this, we employ the following transformations:
\begin{equation}
    \vec{x}=\frac{\vec{r}}{b(t)};\quad \tau \coloneqq \int_0^t \frac{dt}{b^2(t)};\quad \Psi_r(\vec{r},t)=\frac{\psi(\vec{x},t)e^{i\frac{mr^2\partial_t b}{2\hbar b}}}{b},
\end{equation}
as part of the well established scaling approach~\cite{CastinDum,Kagan,2010Gritsev.etalNJP}. However, unlike previous approaches, we relax the assumption that the time-dependence of pairwise interaction potential enters enter via a \textit{single time dependent coupling}, i.e., $V(\vec{r};t)=\mathscr{V}(t)V(\vec{r})$, which in our case would require an isotropic scaling $b_z=b$~\cite{2017Chae.FischerPRL}. Therefore, the dipolar interaction term will in general have an explicit time-dependence in our setup, \textit{besides}
just  the coupling. From this line of argument, we obtain a Lagrangian in the comoving frame of the quasi-2D condensate:
\begin{equation}
    \mathcal{L}_x=\frac{i\hbar}{2}\left(\psi^*\dot{\psi}-\dot{\psi}^*\psi\right)-\frac{\hbar^2}{2m}|\nabla_x\psi|^2-\frac{1}{2}m\omega_0^2f^2x^2|\psi|^2\nonumber-\frac{|\psi_{x}|^2}{2}\int d^2x'V^{\rm 2D}_{\rm int}(\vec{x}-\vec{x}')|\psi_{x'}|^2,
\end{equation}
where derivatives are with respect to the scaling time ($\dot{f}=\partial_{\tau} f$), and in terms of the comoving momentum mode $k$,
\begin{align}
    V^{\rm 2D}_{\rm int}(\vec{x}-\vec{x}')=&\frac{g_c}{\sqrt{2\pi}d_z}\delta^{(2)}(\vec{x}-\vec{x}')+\frac{2g_d}{\sqrt{2\pi}d_z}\int \frac{d^2k}{(2\pi)^2}\left\{1-\frac{3R}{2}w\left[\frac{kd_{z}}{b}\right]\right\}e^{i\vec{k}.(\vec{x}-\vec{x}')},\\
    w[z]=&ze^{\frac{z^2}{2}}\left[1-\erf\left(\frac{z}{\sqrt{2}}\right)\right];\quad f^2=\frac{\omega^2(t)b^4+b^3\partial_t^2b}{\omega_0^2}.\nonumber
\end{align}
In the Madelung representation $\psi=\sqrt{\rho}e^{i\phi}$, the above Lagrangian takes the form:
\begin{equation}
    \mathcal{L}_x=-\hbar \rho\dot{\phi}-\frac{\hbar^2}{8m\rho}(\nabla_x\rho)^2-\frac{\hbar^2\rho}{2m}(\nabla_x\phi)^2-\frac{1}{2}m\omega_0f^2x^2\rho-\frac{1}{2}\int d^2x' V_{\rm int}^{\rm 2D}(\vec{x}-\vec{x}')\rho(x)\rho(x').
\end{equation}
The equations of motion in $\rho$ and $\phi$ are obtained as follows:
\begin{align}
    -\hbar\dot{\rho}&=\frac{\hbar^2}{m}\left[\vec{\nabla_x}\rho.\vec{\nabla_x}\phi+\rho\nabla_x^2\phi\right],\\
    -\hbar\dot{\phi}&=-\frac{\hbar^2}{4m\rho}\nabla_x^2\rho+\frac{\hbar^2}{4m\rho^2}(\nabla_x\rho)^2+\frac{\hbar^2}{2m}(\nabla_x\phi)^2+\frac{1}{2}m\omega_0f^2x^2+\int d^2x' V_{\rm int}^{\rm 2D}(\vec{x}-\vec{x}')\rho(x').
\end{align}
In the comoving frame, the fluid density is approximately constant ($\sim\rho_0$), and the velocity vanishes ($\partial_t\vec{x}=-\frac{\hbar\nabla_x\phi}{mb^2}\sim 0$). Upon linearizing the fluctuations on top of the background density ($\rho_0+\delta\rho$) as well as phase ($\phi_0+\delta\phi$), and neglecting kinetic energy terms ($\nabla_x\rho_0$, $\nabla_x^2\rho_0$) in the Thomas-Fermi approximation, we get:
\begin{align}
    \delta\dot{\phi}+\vec{v}_{\rm com}.\nabla_x\delta\phi&=\frac{\hbar}{4m\rho_0}\nabla^2\delta\rho - \int d^2x' V_{\rm int}^{\rm 2D}(\vec{x}-\vec{x}')\delta \rho(x'),\\
    \delta\dot{\rho}+\vec{v}_{\rm com}.\nabla_x\delta\rho&=-\frac{\hbar}{m}\left(\rho_0\nabla_x^2\delta\phi\right),
\end{align}
where $\vec{v}_{\rm com}=\frac{\hbar}{m}\nabla_x\phi_0$ is the comoving frame velocity. In momentum space, the fluctuations are described in terms of comoving momentum $k$ as follows:
\begin{align}
    \hbar\left(\partial_\tau+i\vec{v}_{\rm com}.\vec{k}\right)\delta \phi_k&=-\left[\frac{\hbar^2k^2}{4m\rho_0}+\frac{g_c}{\sqrt{2\pi}d_z}+\frac{2g_d}{\sqrt{2\pi}d_z}\left\{1-\frac{3R}{2}w\left[\frac{kd_z}{b}\right]\right\}\right]\delta\rho_k\\
     \hbar\left(\partial_\tau+i\vec{v}_{\rm com}.\vec{k}\right)\delta\rho_k&=-\frac{\hbar^2k^2\rho_0}{m}\delta\phi_k
\end{align}
We now synchronize the transverse scaling and coupling strengths as follows, leading to the condition: 
\begin{equation}
    \frac{1}{a^2(t)}=\frac{g_{c}(t)}{g_{c,0}b_z(t)}=\frac{g_d(t)}{g_{d,0}b_z(t)}.
\end{equation}
By setting $\vec{v}_{\rm com}\approx 0$ (negligible comoving frame velocity), we arrive at the following equations of motion:
\begin{align}
    \delta\ddot{\phi}_k&+\left(2\frac{\dot{a}}{a}-\frac{\dot{W}_k}{W_k}\right)\delta\dot{\phi}_k+\frac{c_0^2k^2W_k}{a^2}\delta\phi_k=0,\nonumber\\
    \delta\ddot{\rho}_k&+\frac{c_0^2k^2W_k}{a^2}\delta\rho_k=0,
\end{align}
where we have defined:
\begin{align}
    W_k&=1-\frac{3R}{2}w\left[\frac{b_zkd_{z,0}}{b}\right]+\frac{k^2d_{z,0}^2a^2}{4A};\quad R=\frac{2g_{d,0}}{g_{\rm eff,0}};\quad A=\frac{mc_0^2}{\hbar \omega_{z,0}};\quad  c_0^2=\frac{g_{\rm eff,0}\rho_0}{m}.\nonumber
\end{align}

By further setting the anisotropic scaling condition $b=ab_z$, we get:
\begin{equation}
    W_k=1-\frac{3R}{2}w\left[\frac{kd_{z,0}}{a}\right]+\frac{k^2d_{z,0}^2a^2}{4A},
\end{equation}
{which, away from the free particle regime ($kd_{z,0}a\ll 2\sqrt{A}$) models a modified dispersion $k^2W_k$ equivalent to the functional form $a^2F^2[k\lambda_{pl}/a]$ conjectured in \textit{ad hoc} trans-Planckian models in cosmology, with the cutoff scale $\lambda_{pl}$ being set by the initial transverse trap width $d_{z,0}$:
\begin{equation}
    \frac{a^2}{\lambda_{pl}^2}F^2\left[\frac{k\lambda_{pl}}{a}\right]=\begin{cases} k^2 & \text{Standard}\\
        \frac{a^2}{\lambda_{pl}^2} \tanh ^{2 / p}\left[\left(\frac{k \lambda_{pl}}{a}\right)^p\right] & \text{Unruh}\\
        k^2+k^2 \sum_{q=1}^m \frac{b_q}{(2 \pi)^{2 q}}\left(\frac{\lambda_{pl}}{a}\right)^{2 q} k^{2 q} & \text{Generalized Corley-Jacobson}\\
        k^2-\frac{3Rk^2}{2}w\left[\frac{k\lambda_{pl}}{a}\right] & \text{Proposed BEC analogue; $\lambda_{pl}=d_{z,0}$}
    \end{cases}
\end{equation}
The above functional form ensures that Lorentz-invariance is recovered when modes (of wavelength $\lambda_{\rm phys}=a{k^{-1}}$ in the physical frame) get redshifted beyond the cutoff scale during expansion ($\lambda_{\rm phys}\gg \lambda_{pl}$), and strongly violated when they get blueshifted during contraction ($\lambda_{\rm phys}\ll \lambda_{pl}$). It is also easy to see that $k^2W_k\approx a^2d_{z,0}^{-2}\tanh^{2/p}{\left[(kd_{z,0}/a)^p\right]}|_{p\to1/2}$, wherein at the stability boundary ($g_c=g_d$, or equivalently $R=R_c$), the BEC analogue almost exactly matches with the Unruh dispersion. The evolution of the modified dispersion $kW_k^{1/2}$ can be observed in \ref{fig:Dispersion}.

\begin{figure}[hbt]
\centering
\includegraphics[scale=0.5]{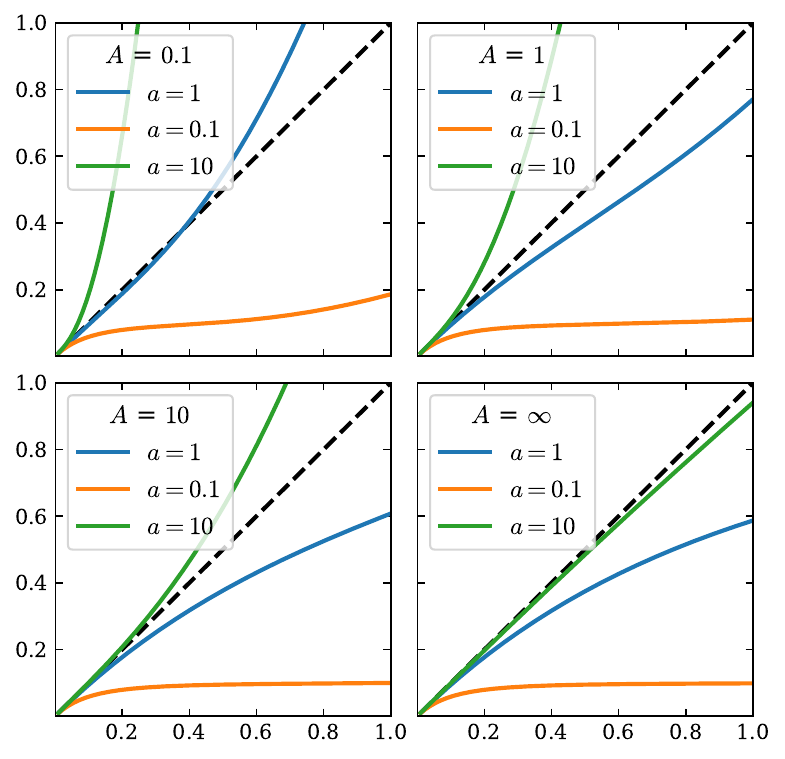}
\caption{Modified dispersion $kW_k^{1/2}$ (y-axis) with respect to wavenumber $k$ (x-axis) for fixed dipolar strength $R=R_c$ (Unruh value) and various values of the scale factor $a$ and dimensionless sound speed $A$. The dashed black line corresponds to the standard Lorentz-invariant dispersion. On suppressing the free-particle term completely ($A\to\infty$), the dispersion exactly captures cosmological transplanckian dynamics wherein Lorentz-violation dominates the early stages of expansion ($a\sim1$) and the late stages of contraction ($a\ll1$). For small values of $A$, the free-particle term dominates large-$k$ modes as expansion proceeds ($a>1$) or in the early-stages of collapse ($a\sim 1$).}
\label{fig:Dispersion}
\end{figure}

In this setup, the comoving frame of the fluid exactly matches the comoving frame of the analogue space-time.} However, the lab frame ($\equiv b_z\vec{x}$) and the analogue physical frame ($\equiv a\vec{x}$) can be different depending on our choice of implementing the synchronization conditions:
\begin{equation}
    \frac{b_z}{a^2}=\frac{g_{c}}{g_{c,0}}=\frac{g_d}{g_{d,0}}\quad\&\quad b=ab_z,
\end{equation}
which are achieved by tuning the transverse and radial trapping frequencies as follows:
\begin{equation}
    \omega_z(t)=\sqrt{\frac{\omega_{z,0}^2}{b_z^4}-\frac{\partial_t^2b_z}{b_z}};\quad \omega(t)=\sqrt{\frac{\omega_{0}^2f^2}{b^4}-\frac{\partial_t^2b}{b}}.
\end{equation}
For the isotropic scaling case, setting $f^2=a^{-2}$ is sufficient to obtain a similar scaling relation for the phase as $\phi_0(\tau)=-\hbar^{-1}\mu_0\int  f^2 d\tau$~\cite{2017Chae.FischerPRL}. However this does not carry over to the anisotropic scaling setup, for which we will have to rely on numerical simulations. 
The conformal time $\eta$ used in the analysis is related to the scaling time $\tau$ and the lab time $t$ as follows:
\begin{equation}
    d\eta=\frac{d\tau}{a}=\frac{dt}{ab^2} \implies \eta=\int\frac{d\tau}{a}=\int\frac{dt}{ab^2}.
\end{equation}

We discuss two main ways in which the Planck-scale sensitive cosmological quantum simulator can be implemented:
\begin{itemize}
    \item \textbf{Time-independent transverse trap:} The gas radially expands with the same scale factor as the analogue-cosmological background, i.e., the lab frame and physical frame coincide:
    \begin{equation}
        b=a;\quad b_z=1;\quad \frac{g_{c}}{g_{c,0}}=\frac{g_d}{g_{d,0}}=\frac{1}{a^2}.
    \end{equation}
    For a power-law model $a=(\eta/\eta_i)^v$ in conformal time, the trapping frequencies take the form:
    \begin{equation}
        \omega_z^2(\eta)=\omega_{z,0}^2;\quad \omega^2(\eta)=\frac{\omega_{0}^2}{a^4}+\frac{v(2v+1)}{a^2\eta^2}\quad \text{where}\quad \frac{\eta}{\eta_i}=\left(\frac{t}{t_i}\right)^{\frac{1}{3v+1}},
    \end{equation}
    For both $v=-1$ and $v=3$, we see that the above trapping frequencies remain real throughout the evolution i.e., $\omega_z^2,\omega^2>0$.
    \item \textbf{Time-independent coupling strengths:} Here, we do not change the coupling strengths, however the gas must expand appropriately to maintain synchronization, i.e., the lab frame and physical frame do not coincide:
    \begin{equation}
        b=a^3;\quad b_z=a^2;\quad \frac{g_{c}}{g_{c,0}}=\frac{g_d}{g_{d,0}}=1.
    \end{equation} 
    In this case, the trapping frequencies take the form:
    \begin{equation}
        \omega_z^2(\eta)=\frac{\omega_{z,0}^2}{a^8}+\frac{2v(5v+1)}{a^{14}\eta^2};\quad \omega^2(\eta)=\frac{\omega_{0}^2}{a^{12}}+\frac{3v(4v+1)}{a^{14}\eta^2}\quad \text{where}\quad \frac{\eta}{\eta_i}=\left(\frac{t}{t_i}\right)^{\frac{1}{7v+1}},
    \end{equation}
    For both $v=-1$ and $v=3$, we see that the above trapping frequencies remain real throughout the evolution, i.e., $\omega_z^2,\omega^2>0$.
\end{itemize}
\section{Validity regime of cosmological Planck-scale effects}
\label{Sec. II}

\begin{figure}[hbt]
\centering
\includegraphics[scale=0.5]{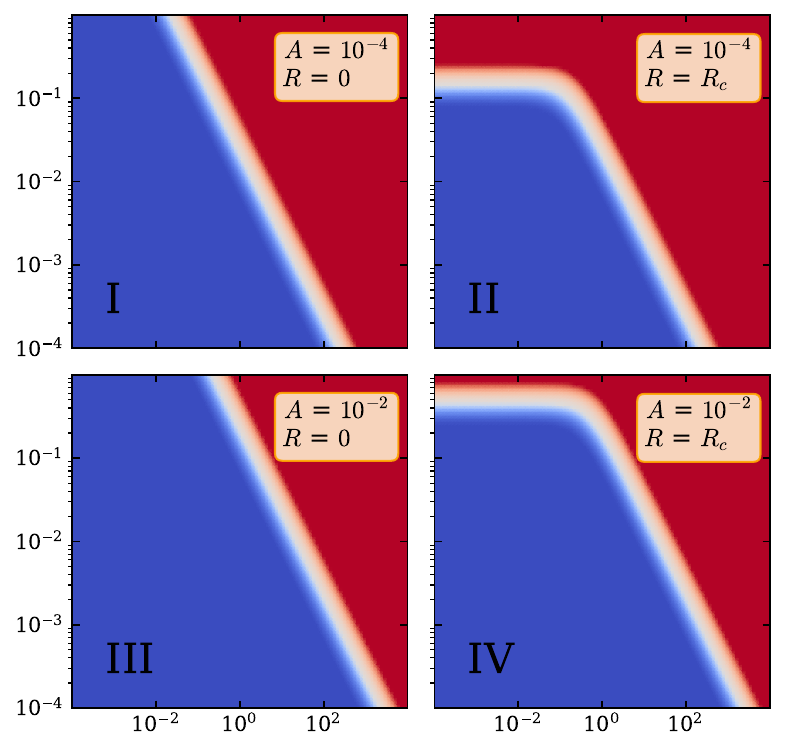}
\caption{Validity regime for the analogue cosmology mapping in dipolar condensates in the parameter space of scale factor and  momentum, i.e., $(a,k)$ upon setting $d_{z,0}=1$. The blue region ($\gamma<0.1$ in the figure) is enhanced in the low-momentum sector, thus delaying the crossover to the red region ($\gamma>10$) during expansion. For contraction, the simulation progresses deeper into the analogue-cosmology regime (blue), thereby avoiding the free-particle crossover. Although larger values of $A$ can further enhance the blue region, the limit $A\gg1$ is at odds with the assumption that the transverse trap-width and the oscillator width are related exactly via the Ermakov equation.}
\label{fig:ValidityRegime}
\end{figure}

While the third (free particle) term in $W_k$ resembles a Corley-Jacobson type modification (\ref{fig:disp}), its time-dependence ($\propto a^2$) causes an amplification of Lorentz-violation with the redshifting of modes during expansion, directly contradicting the trans-Planckian features of cosmological models. It is therefore ideal for both the Lorentz-invariant as well as the trans-Planckian regimes to be probed well before free particles take over (after which an exact cosmology mapping begins to break down). While the limit $k^2d_{z,0}^2a^2\ll 4A$ used in the main text effectively captures the ``pure cosmology" regime, a more precise quantification is possible via the following parameter:
    \begin{equation}
        \gamma \coloneqq\frac{W_k|_{\rm FP}}{W_k|_{\rm TP}};\quad W_k|_{\rm FP}=\frac{k^2d_{z,0}^2a^2}{4A};\quad W_k|_{\rm TP}=1-\frac{3R}{2}w\left[\frac{kd_{z,0}}{a}\right]
    \end{equation}
which when small ($\gamma \ll1$) preserves the cosmology mapping, and when large ($\gamma\gg1$) indicates that free particles have taken over (\ref{fig:ValidityRegime}). Since the high-momentum imprints of Planck-scale physics may be drastically affected by the free-particles in the simulator, we focus on low-momentum imprints with direct cosmological implications as discussed in the main text.

Note that however the quasi-2D limit taken for the transverse trap width to be determined by the standard Ermakov equation~\eqref{eq:Ermakov1} corresponds to $A\ll1$, which can potentially be in conflict with that of $\gamma\ll1$ corresponding to a ``pure cosmology" regime. For instance, engineering the Unruh dispersion at the beginning stages of expansion is only possible when $A\gg 1$ (the mapping also gets progressively worse as the expansion proceeds). In this regime, the current ansatz serves only as a crude approximation, whereas mean-field effects from the transverse dimension must be taken into account for a more faithful treatment. However, contrary to expanding models, Planck-scale effects are anticipated at the \textit{end stages} of contraction ($a\ll1$), where we see that the free particle term vanishes to give rise to the Unruh dispersion for a wide-range of $k$-values even when $A\ll1$ (see \ref{fig:Dispersion}).

As discussed earlier, to preserve the analogue cosmology mapping exactly, it is ideal to confine to low-momentum modes where the free particle term can be safely dropped ($k^2d_{z,0}^2a^2\ll 4A$). This is compatible with the quasi-2D ansatz ($A\ll 1$) for long enough time scales provided the initial trap-width $d_{z,0}$ is sufficiently small ($\omega_{z,0}\gg \omega_0$). However, away from this regime, the free-particle contribution must be included, which can lead to additional corrections exclusive to the analogue setup. These corrections better address the complementary regime of $k^2d_{z,0}^2a^2\gtrsim 4A$, where the results from \eqref{eq:correction} no longer apply. Staying within the quasi-2D regime ($A\ll 1$), this leads to a modified late time ($\eta\to0^-$) dispersion as follows:
     \begin{equation}\label{eq:correction2}
       \omega_k^2 \simeq 
        \begin{cases}  k^2W_k+\frac{3}{4\eta^2}\left(\frac{1}{3}+\frac{32A}{3 k^2 a^2}\right) &  v=-1\\ k^2W_k-\frac{3}{4\eta^2}\left(5+\frac{32 A (3 a^4 R + 5 k^4 (R - Rc))}{a^2 k^6 R_c}\right) &  v=3 \end{cases},
    \end{equation}
where we observe some interesting consequences. First, the damping term $\Delta_k$ is promoted to leading order at late-times similar to the critical ``Unruh" case discussed in \eqref{eq:critdisp}. For inflation, this causes the superhorizon power spectrum to mimic the features of a $v=1$ radiation dominated expansion, whereas for contraction it mimics the features of a $v=-3$ expansion (same as \eqref{eq:critdisp}). Second, even after incorporating the free-particle term, both duality breaking as well as the contraction signatures of damping are preserved in the experimental setup.

\section{Generality of duality breaking and UV/IR imprints in cosmology}
\label{Sec. III} 
{In this section we show that the analysis of duality breaking presented in the main text is not tied to the analogue system but more generally applicable to cosmological trans-Planckian models. Let us assume a more general form for the damping terms entering the effective frequency as follows (with $W_k$ expanded upto subleading order):
\begin{equation} 
\omega_k^2=k^2W_k+\frac{v(2-v)}{4\eta^2}+\frac{A_1W_k'^2}{W_k^2}+\frac{A_2W_k''}{W_k}+\frac{A_3a'W_k'}{aW_k};\quad W_k\sim \begin{cases}
1+c_l(k/a)^l & k/a\to 0 \,\,\,\,\,\text{($l>0$)}\\
c_{m_1}(k/a)^{m_1}+c_{m_2}(k/a)^{m_2} &k/a \to \infty \text{  ($m_1>m_2$)}
\end{cases}\nonumber
\end{equation}
which (i) satisfies dimensional considerations ($\propto \eta^{-2}$), (ii) recovers Lorentz-invariance for long-wavelength modes ($k\to 0$), and (iii) recovers duality ($v\to 2-v$) for a static trans-Planckian modification $W_k$. Note that ad hoc models previously conjectured do not take into account such corrections, i.e., they assume $A_i=0$. With some algebra, the above expression can be rewritten as follows,:
\begin{equation}\label{eq:general1}
    \omega_k^2=k^2W_k+\frac{v(2-v)}{4\eta^2}\left(1+\Delta_k\right); \quad \Delta_k=B_1\frac{a\partial_aW_k}{W_k}+B_2\left(\frac{a\partial_aW_k}{W_k}\right)^2+B_3\frac{a^2\partial_a^2W_k}{W_k},
\end{equation}
from which the analogue model in \eqref{eq:disp} can be recovered as a special case. For late-time limits corresponding to redshifting ($k/a\ll1$) or blueshifting ($k/a\gg1$) modes, the terms in $\Delta_k$ asymptote to:
\begin{align}
    \frac{a\partial_aW_k}{W_k}&\sim \begin{cases}
-lc_l(k/a)^l  & k/a\to 0\\
-m_1+\frac{(m_1-m_2)c_{m_2}}{c_{m_1}}(a/k)^{m_1-m_2} &k/a \to \infty 
\end{cases}\\
\frac{a^2\partial_a^2W_k}{W_k}&\sim \begin{cases}
l(l+1)c_l(k/a)^l  & k/a\to 0\\
m_1(m_1+1)+\frac{\left[m_2(m_2+1)-m_1(m_1+1)\right]c_{m_2}}{c_{m_1}}(a/k)^{m_1-m_2} &k/a \to \infty
\end{cases},
\end{align}
which results in a general asymptotic form for $\Delta_k$ as follows:
\begin{equation}\label{eq:gendelta}
    \Delta_k\sim \begin{cases}
C_0(k/a)^l  & k/a\to 0\\
D_0m_1+D_1m_1^2+D_3(a/k)^{m_1-m_2} &k/a \to \infty \text{  ($m_1>m_2$)}
\end{cases},
\end{equation}
thereby confirming that nonadiabatic corrections are subleading for $m_1=0$ (e.g., noncritical case $R<R_c$), dominated by high-k modes during inflation and low-k modes during contraction. For $m_1\neq 0$ (such as in the Unruh case $R=R_c$) the correction is promoted to leading order for contraction, dramatically affecting the mode evolution (for e.g., the power spectrum freezes instead of amplifying when $R=R_c$). The results obtained in this work are therefore not tied to the analogue, but more generally applicable to cosmological models that incorporate nonadiabatic corrections arising from trans-Planckian physics. Some special cases are obtained as follows:
\begin{align}
    \text{Corley-Jacobson : } l=2,\,m_1=2,\, m_2=0 &\implies \Delta_k\propto\begin{cases}
(k/a)^2  & k/a\to 0\\
1+\tilde{D}_3(a/k)^{2} &k/a \to \infty
\end{cases}\nonumber\\
    \text{BEC analogue ($R<R_c$) : } l=1,\,m_1=0,\,m_2=-2 &\implies \Delta_k\propto\begin{cases}
k/a  & \quad\quad\quad k/a\to 0\\
(a/k)^{2} &\quad\quad\quad k/a \to \infty
\end{cases} \nonumber\\
    \text{BEC (Unruh case: $R=R_c$) : } l=1,\,m_1=-2,\,m_2=-4 &\implies \Delta_k\propto\begin{cases}
k/a  & k/a\to 0\\
1+\tilde{D}_3(a/k)^{2} &k/a \to \infty
\end{cases} 
\end{align}
We also show that this analysis holds up for a general minimal length cosmological model in 3+1-dimensions, wherein modified commutation relations lead to a new conjugate variable in place of physical momenta, i.e., $k/a\to k^2W_k^{1/2}/a$~\cite{2003Hassan.SlothNPB}:
\begin{equation}
    \omega_k^2=k^2W_k-\frac{\partial_\eta^2\left(a\sqrt{J_k}\right)}{a\sqrt{J_k}};\quad J_k=\frac{\partial \left(k^3W_k^{3/2}/a^3\right)}{\partial(k^3/a^3)}=W_k^{3/2}+\left(\frac{k\sqrt{W_k}}{2a}\right)\frac{\partial W_k}{\partial(k/a)} 
\end{equation}
resulting in a similar form as in \eqref{eq:general1}, but with $W_k$ replaced by $J_k$ (i.e., the Jacobian from transforming the momenta as $k/a\to kW_k^{1/2}/a$) in the nonadiabatic correction $\Delta_k$:
\begin{align} 
\omega_k^2=k^2W_k+\frac{v(1-v)}{\eta^2}+\frac{A_1J_k'^2}{J_k^2}+\frac{A_2J_k''}{J_k}+\frac{A_3a'J_k'}{aJ_k}&\implies \omega_k^2=k^2W_k+\frac{v(1-v)}{\eta^2}\left(1+\Delta_k\right)\\
W_k\sim \begin{cases}
1+c_l(k/a)^l & k/a\to 0 \,\,\,\text{  ($l>0$)}\\
c_{m_1}(k/a)^{m_1}+c_{m_2}(k/a)^{m_2} &k/a \to \infty \text{  ($m_1>m_2$)}
\end{cases}&\implies J_k\sim \begin{cases}
1+\tilde{c}_l(k/a)^l & k/a\to 0\\
\tilde{c}_{m_1}(k/a)^{\frac{3m_1}{2}}+\tilde{c}_{m_2}(k/a)^{m_2+\frac{m_1}{2}} &k/a \to \infty
\end{cases} \nonumber
\end{align}
Note that the duality condition above is with respect to 3+1-dimensions~\cite{1999WandsPRD}, i.e., $v\to 1-v$. The co-efficients are still kept arbitrary so as to carry over the results to any space-time dimensions. Repeating the same steps as before, we obtain the exact same general asymptotic form for $\Delta_k$ as in \eqref{eq:gendelta}:
\begin{equation}
    \Delta_k\sim \begin{cases}
C_0(k/a)^l  & k/a\to 0\\
D_0m_1+D_1m_1^2+D_3(a/k)^{m_1-m_2} &k/a \to \infty \text{  ($-2\geq m_1>m_2$)}
\end{cases}
\end{equation}
The only additional constraint here is that unbounded modifications are ruled out by the minimal length principle, i.e., $k^2W_k/a^2\leq \lambda_{pl}^{-2}$, hence naturally giving rise to Unruh-like dispersions discussed in the main text. This imposes $m_1\leq-2$ for the asymptotic expansion for $W_k$, due to which the nonadiabatic correction $\Delta_k$ is generally promoted to leading order during contraction for minimal length models  (similar to the $R=R_c$ case in the BEC analogue). From these considerations, we argue that duality breaking and UV/IR signatures of trans-Planckian damping are quite robust to the underlying theory.}

\section{Power spectrum simulation}\label{Sec. IV}
Suppose the vacuum is prepared at some finite-time $\eta_i$, each mode evolves as a harmonic oscillator with a time dependent frequency, while retaining its Gaussian form~\cite{2008LoheJPA,2024Chandran.etalPRD}:
\begin{equation}
    \Psi_k(\delta\hat{\bar{\phi}}_k,\eta)=\left(\frac{\omega_k^{\rm in}}{\pi b_k^2}\right)^{1/4}\exp\left[-\left(\frac{\omega_k^{\rm in}}{b_k^2}-\frac{ib'_k}{b_k}\right)\frac{\delta\hat{\bar{\phi}}_k^2}{2}-\frac{i\omega_k^{\rm in}}{2}\int\frac{d\eta}{b_k^2}\right];\quad \omega_k^{\rm in}=\omega_k(\eta_i),
\end{equation}
where the scaling parameters $\{b_k\}$ are the solutions to the nonlinear Ermakov-Pinney equation~\cite{1950PinneyPotAMS}, 
\begin{equation}\label{eq:ermakov}
    b''_k(\eta)+\omega_k^2(\eta)b(\eta)=\frac{(\omega_k^{\rm in})^2}{b_k^3(\eta)}, 
\end{equation}
that satisfy the initial conditions $b_k(\eta_i)=1$ and $b_k'(\eta_i)=0$. Note that the above scaling parameters are different from the parameters $b(t)$ and $b_z(t)$ employed in the scaling approach for BEC. The Ermakov-Pinney scaling parameters and mode-functions $\delta\bar{\phi}_k$ are related as follows:
\begin{equation}
    |\delta\bar{\phi}_k|^2=\langle\delta\hat{\bar{\phi}}_k\delta\hat{\bar{\phi}}_k\rangle=\frac{b_k^2}{2\omega_k^{\rm in}},
\end{equation}
where the mode functions evolve corresponding to the frequencies in \eqref{eq:disp}, from a vacuum-state defined at $\eta=\eta_i$:
\begin{equation}
    \delta\bar{\phi}''_k+\omega_k^2(\eta)\delta\bar{\phi}_k=0;\quad \delta\bar{\phi}_k(\eta_i)=\frac{e^{-i\omega_k^{\rm in}\eta_i}}{\sqrt{2\omega_k^{\rm in}}}.
\end{equation}
Finally, the power spectrum can be obtained from the numerical solutions of \eqref{eq:ermakov} as follows:
\begin{equation}
    \mathcal{P}_{\delta\phi}=k^2|\delta\phi_k|^2=\frac{k^2W_k}{a}|\delta\bar{\phi}_k|^2=\frac{k^2W_kb_k^2}{2a\omega_k^{\rm in}}.
\end{equation}
Note that confining to positive frequency modes $\omega_k^{\rm in}>0$ at a \textit{finite} initial time places a lower bound on the wavenumber. This also requires us to avoid the supercritical regime of dipolar strength ($R>R_c$) which can lead to a deep roton minimum at large $k$. We therefore avoid the zero-mode/inverted-mode imprints that occur due to nonstandard initial states, and extract Planck-scale effects that exclusively arise in a stable, minimum-energy vacuum state. For \ref{fig:PST} and \ref{fig:PS2} in the main text, the initial time is therefore set to be $\eta_i=-10^{-3}$ to probe low enough momentum modes.

\end{widetext}
\end{appendix}




\bibliography{pss_v12}

\end{document}